\documentclass[twocolumn,showpacs,preprintnumbers,amsmath,amssymb]{revtex4}

\usepackage{graphicx}
\usepackage{dcolumn}
\usepackage{bm}

\begin{document} 

\preprint{ }

\title{Ferromagnetism and Electronic Structures of Nonstoichiometric Heusler-Alloy Fe$_{3-x}$Mn$_{x}$Si Epilayers Grown on Ge(111)}

\author{K. Hamaya,$^{1,2}$\footnote{E-mail: hamaya@ed.kyushu-u.ac.jp} H. Itoh,$^{3}$ O. Nakatsuka,$^{4}$ K. Ueda,$^{1}$ K. Yamamoto,$^{1}$ M. Itakura,$^{5}$\\ T. Taniyama,$^{2,6}$ T. Ono,$^{7}$ and M. Miyao$^{1}$\footnote{E-mail: miyao@ed.kyushu-u.ac.jp}}
\affiliation{%
$^{1}$Department of Electronics, Kyushu University, 744 Motooka, Fukuoka 819-0395, Japan}%
\affiliation{%
$^{2}$PRESTO, Japan Science and Technology Agency, 4-1-8 Honcho, Kawaguchi 332-0012, Japan}%
\affiliation{%
$^{3}$Department of Pure and Applied Physics, Kansai University, Suita 564-8680, Japan}%
\affiliation{%
$^{4}$Department of Crystalline Materials Science, Nagoya University, Chikusa-ku, Nagoya 464-8603, Japan}%
\affiliation{%
$^{5}$ Department of Applied Science for Electronics and Materials, Kyushu University, 6-1 Kasuga, Fukuoka 816-8580, Japan}
\affiliation{%
$^{6}$Materials and Structures Laboratory, Tokyo Institute of Technology, 4259 Nagatsuta, Yokohama 226-8503, Japan}%
\affiliation{%
$^{7}$Institute for Chemical Research, Kyoto University, Uij, Kyoto 611-0011, Japan}%

%

\date{\today}
\begin{abstract}
For the study of ferromagnetic materials which are compatible with group-IV semiconductor spintronics, we demonstrate control of the ferromagnetic properties of Heusler-alloys Fe$_{3-x}$Mn$_{x}$Si epitaxially grown on Ge(111) by tuning the Mn composition $x$. Interestingly, we obtain $L2_\text{1}$-ordered structures even for nonstoichiometric atomic compositions. The Curie temperature of the epilayers with $x \approx$ 0.6 exceeds 300 K. Theoretical calculations indicate that the electronic structures of the nonstoichiometric Fe$_{3-x}$Mn$_{x}$Si alloys become half-metallic for 0.75 $\le$ $x$ $\le$ 1.5. We discuss the possibility of room-temperature ferromagnetic Fe$_{3-x}$Mn$_{x}$Si/Ge epilayers with high spin polarization. 
\end{abstract}
\maketitle
By introducing spin degrees of freedom into group-IV semiconductor-based electronic devices, it becomes possible to add novel functions to existing silicon large-scale integrated circuit (LSI) technologies.\cite{Min,Ian,Jonker1,Jonker2,Nakane} Group-IV semiconductor spintronics also enables us to overcome the scaling limits of silicon-based complementary metal--oxide--semiconductor devices. To realize highly efficient spin injection and detection in group-IV semiconductor devices, it is crucial to develop compatible ferromagnetic materials, which can be grown epitaxially on Si and/or Ge, with high spin polarization and high Curie temperature. In this context, we have focused on ferromagnetic full-Heusler alloys with the chemical formula  $X$$_{2}$$Y$$Z$, where $X$ and $Y$ are transition metals and $Z$ is a main group element such as Si and Ge. In general, full-Heusler alloys become half-metallic ferromagnets (HMFs) with a fully spin-polarized density of states (DOS) at the Fermi level (100\% spin polarization).\cite{Groot,Lu,Sakuraba,Inomata} 

Recently, we demonstrated highly epitaxial growth of ferromagnetic Heusler-type alloys, including Fe$_{3}$Si and Fe$_{2}$MnSi thin films, on the group-IV semiconductors Si and Ge, using low-temperature molecular beam epitaxy (LT-MBE).\cite{Sadoh,Ueda,Hamaya} Note that the interface between these ferromagnets and Si or Ge has atomic-scale abruptness and an ordered structure can be obtained in spite of low-temperature growth at 130 and 200 $^{\circ}$C.\cite{Sadoh,Ueda,Hamaya} Since Fe$_{3}$Si has a high Curie temperature above 800 K, we can expect that spin devices made with this material will exhibit room-temperature operation. For this material, the highest reported spin polarization to date is $|P|$ $\sim$ 45 $\pm$ 5\%.\cite{Ionescu} Fe$_{2}$MnSi is predicted to be an HMF,\cite{Fujii,Hongzhi} and it is anticipated that this material can be successfully applied to highly efficient spin injection and detection through Schottky tunnel barriers in group-IV semiconductor devices. However, epitaxial Fe$_{2}$MnSi thin films have a Curie temperature of $\sim$ 210 K, which is much lower than room temperature.\cite{Ueda} To realize group-IV semiconductor spintronics, we require a convenient ferromagnetic material that simultaneously exhibits the advantages of the above two characteristics, namely an HMF with a high Curie temperature. 

The study of bulk Fe$_{3-x}$Mn$_{x}$Si, reported by Yoon and Booth,\cite{Yoon} shows that the magnetic properties can be tuned by controlling the Mn composition $x$. Bulk samples with $x \le$ 0.85 yield room-temperature ferromagnetism, although we note that the samples were fabricated only by high-temperature annealing (800 $^{\circ}$C) and rapid quenching. In this Letter, we focus on {\it epitaxial} Fe$_{3-x}$Mn$_{x}$Si layers grown on a group-IV semiconductor Ge for use in semiconductor-based spintronic applications. We demonstrate control of the ferromagnetic properties of Fe$_{3-x}$Mn$_{x}$Si and achieve room-temperature ferromagnetic epilayers with an ordered $L2_\text{1}$ structure even for nonstoichiometric atomic compositions. Theoretical calculations suggest that Fe$_{3-x}$Mn$_{x}$Si alloys with compositions in the range 0.75 $\le$ $x$ $\le$ 1.5 become HMFs. We confirm that a high spin polarization ($P \ge$ 0.9) can be achieved for compositions in the range 0.5 $\le$ $x$ $<$ 0.75. We also discuss the possibility of room-temperature ferromagnetic Fe$_{3-x}$Mn$_{x}$Si/Ge epilayers with a high spin polarization. 

Fe$_{3-x}$Mn$_{x}$Si layers with a thickness of $\sim$ 50 or $\sim$ 100 nm were grown on n-type Ge (111) by molecular beam epitaxy (MBE). We employed a surface cleaning process described in a previous work.\cite{Ueda} Prior to the growth of the Fe$_{3-x}$Mn$_{x}$Si layers, 30-nm-thick Ge buffer layers were grown at 400 $^{\circ}$C with a growth rate of 0.60 nm/min. After confirming streak patterns by {\it in-situ} reflection high energy electron diffraction (RHEED), the substrate temperature was reduced to 200 $^{\circ}$C. Using Knudsen cells, we co-evaporated Fe, Mn, and Si. In order to change the Mn composition $x$, the growth rate of Mn was tuned by adjusting the cell temperature. After the growth, we observed RHEED patterns of the Fe$_{3-x}$Mn$_{x}$Si layers for various $x$, as shown in Fig. 1(a). We determined $x$ by energy dispersive x-ray spectroscopy and Rutherford backscattering spectroscopy measurements. All the RHEED patterns clearly show symmetrical streaks, indicating good two-dimensional epitaxial growth of the Fe$_{3-x}$Mn$_{x}$Si layers on Ge(111). We note that even for nonstoichiometric atomic compositions, epitaxial growth is indicated.
\begin{figure}[t]
\includegraphics[width=8.5cm]{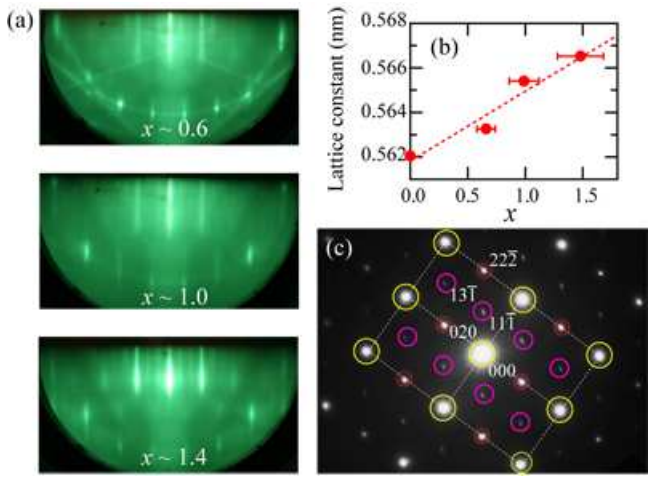}
\caption{(Color online) (a) Reflection high energy electron diffraction (RHEED) patterns observed along the [$\bar{2}$11] azimuth for Fe$_{3-x}$Mn$_{x}$Si layers grown at 200$^{\circ}$C with $x \approx$ 0.6, 1.0, and 1.4. (b) Lattice constant calculated from $d_{224}$ vs. $x$. (c) Cross-sectional selected-area diffraction patterns for $x \approx$ 0.6. The zone axis is parallel to the [101] direction.}
\end{figure} 
 
For the 100-nm-thick epilayers, we performed structural characterization by high-resolution x-ray diffraction with Cu$K_{\alpha}$ radiation. As described in Ref. 11, since the (111) diffraction peak of the Fe$_{3-x}$Mn$_{x}$Si layers could not be separated from the Ge(111) peak in $\theta$-2$\theta$ scans, we could not measure the out-of-plane lattice constant. To separate the diffraction peaks of the Ge substrate from those of the Fe$_{3-x}$Mn$_{x}$Si layers, we measured reciprocal space maps for Fe$_{3-x}$Mn$_{x}$Si(224), together with Ge (331). Assuming a cubic crystal structure, we can estimate the lattice constant for various $x$ from the (224) spacing $d_{224}$. A plot of lattice constant (calculated from $d_{224}$) vs. $x$ is displayed in Fig. 1(b), together with that of an epitaxial Fe$_{3}$Si ($x =$ 0) layer grown on Ge(111).\cite{Sadoh} The figure shows that the lattice constant lengthens almost linearly with increasing $x$. This tendency is in good agreement with that of bulk samples in previous works.\cite{Yoon,Niculescu} This result indicates that the doped Mn is systematically tuned. We also present cross-sectional selected-area diffraction (SAD) patterns of an Fe$_{3-x}$Mn$_{x}$Si epilayer with $x \approx$ 0.6 in Fig. 1(c). The SAD patterns clearly show super-lattice reflections due to the presence of the ordered $L2_\text{1}$ structure, indicated by the pink solid circles. The two other diffraction patterns express the fundamental and super-lattice reflections, corresponding to the $A2 + B2 + L2_\text{1}$ (yellow) and the ordered $B2 + L2_\text{1}$ structures (red), respectively. Namely, the Fe$_{3-x}$Mn$_{x}$Si epilayers include the ordered $L2_\text{1}$ structure even for nonstoichiometric atomic compositions.
\begin{figure}[t]
\includegraphics[width=9cm]{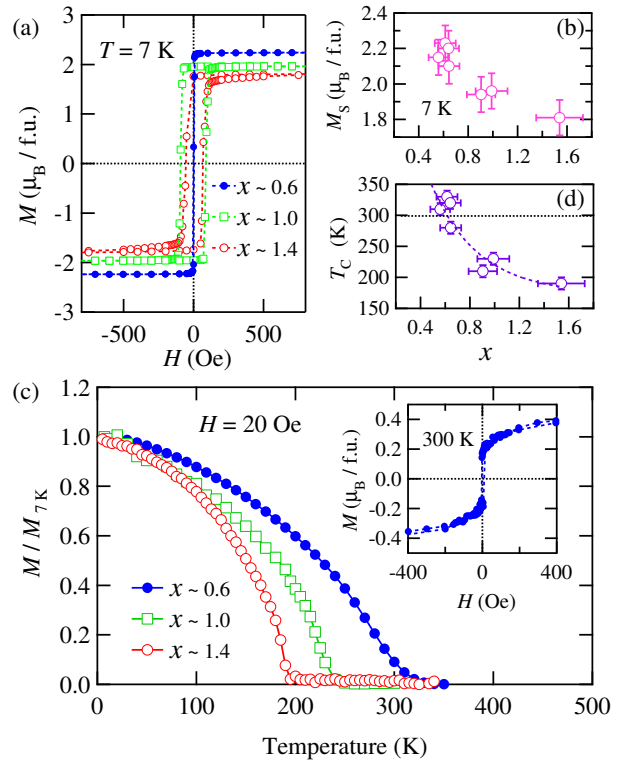}
\caption{(Color online) (a) $M$--$H$ curves of Fe$_{3-x}$Mn$_{x}$Si epilayers for various $x$, measured at 7 K. (b) $M_\text{s}$ vs. $x$ at 7 K. (c) Temperature dependence of the normalized magnetization for three samples with $x \approx$ 0.6, 1.0, and 1.4. The inset shows the $M$--$H$ curve for $x \approx$ 0.6, measured at 300 K. (d) $T_\text{C}$ vs. $x$.}
\end{figure}  

Using a superconducting quantum interference device (SQUID) magnetometer, we measured the magnetic properties of Fe$_{3-x}$Mn$_{x}$Si epilayers for various $x$. Figure 2 (a) shows the field-dependent magnetization ($M$--$H$) for $x \approx$ 0.6, 1.0, and 1.4 at 7 K, where the vertical axis expresses the magnetic moment per formula unit. The magnetic field is applied parallel to the film plane with crystal orientation [0$\overline{1}$1]. All the epilayers exhibit clear ferromagnetic hysteretic curves and the shape of the $M - H$ curves varies slightly for changes in $x$. For $x \approx$ 0.6, the coercivity is quite small compared to that of Fe$_{2}$MnSi/Ge layers (i.e., $x =$ 1)\cite{Ueda} and is similar to that of Fe$_{3}$Si/Ge layers.\cite{Sadoh,Ionescu,Ploog} The saturation magnetization ($M_\text{s}$) at 7 K is summarized for various $x$ in Fig. 2(b). $M_\text{s}$ gradually changes with doping Mn composition.\cite{Yoon} Figure 2(c) shows the temperature-dependent magnetization ($M - T$) for various $x$ in a small magnetic field of 20 Oe, where the magnetization is normalized by that at 7 K ($M$/$M_\text{7 K}$). For $x \approx$ 1.0 (nearly stoichiometric composition), $M$/$M_\text{7 K}$ disappears at $\sim$ 230 K, i.e., a Curie temperature ($T_\text{C}$) $=$ 230 K, largely consistent with that of bulk Fe$_{2}$MnSi samples.\cite{Hongzhi,Zhang} For a nonstoichiometric composition of $x \approx$ 0.6 or 1.4, $T_\text{C}$ becomes higher or lower than 230 K, respectively. It should be noted that for $x \approx$ 0.6, finite magnetic moments are observed at 300 K. We also measured the $M$--$H$ curve at 300 K, as shown in the inset of Fig. 2(c), and an evident ferromagnetic feature can be seen, i.e., room-temperature ferromagnetism. Figure 2(d) displays $T_\text{C}$ as a function of $x$. $T_\text{C}$ also systematically changes with varying $x$ and room-temperature ferromagnetism is obtained at $x$ $\approx$ 0.6. These results demonstrate that our LT-MBE technique can realize control of ferromagnetism of Fe$_{3-x}$Mn$_{x}$Si epilayers. We emphasize that the room-temperature ferromagnetic Fe$_{3-x}$Mn$_{x}$Si epilayers include an $L2_\text{1}$-ordered structure, as described in Fig. 1(c).

In order to discuss the electronic structure of the examined Fe$_{3-x}$Mn$_{x}$Si with nonstoichiometric atomic compositions ($0 \le x \le 1.5$), we carried out first-principles band calculations using the Vienna {\it ab initio} simulation package (VASP).\cite{Kresse} The calculations are based on density functional theory (DFT) in the generalized gradient approximation (GGA). Although the electronic structure of Fe$_{3-x}$Mn$_{x}$Si has been studied for $0 \le x \le 0.5$\cite{Go} using the TB-LMTO-ASA method,\cite{Andersen} the electronic structure for $x > 0.5$ has not yet been clarified. We hereafter consider that the crystal lattice structures of Fe$_3$Si and Fe$_2$MnSi are $DO_\text{3}$ and $L2_\text{1}$ types, respectively, and that the unit cell is composed of four interpenetrating fcc sub-lattices originating at A: $(0, 0, 0)$, B: $(\frac{1}{4},\frac{1}{4},\frac{1}{4})$, C: $(\frac{1}{2},\frac{1}{2},\frac{1}{2})$, and D: $(\frac{3}{4},\frac{3}{4},\frac{3}{4})$. For both Fe$_3$Si and Fe$_2$MnSi, (A, C) sites and D sites are occupied by Fe atoms and Si atoms, respectively, and B sites are occupied by Fe atoms for Fe$_3$Si and by Mn atoms for Fe$_2$MnSi. For nonstoichiometric compositions, we consider super-cells consisting of eight atoms ($x = 0.5$), 16 atoms ($x = 0.25$, $0.75$, $1.25$, and $1.5$), 32 atoms ($x = 0.125$ and $0.875$), and 128 atoms ($x = 0.625$). For $0 < x \le 1$, we assume that Mn atoms enter B sites, as indicated by previous experiments.\cite{Yoon} For $1 < x \le 1.5$, we assume that all B sites and some (A, C) sites are occupied by Mn atoms. Figures 3(a) and (b) show the densities of states (DOS), $D_\sigma(E)$, calculated for spin-up ($\sigma = \uparrow$) and spin-down ($\sigma = \downarrow$) states, respectively. Here, we used a lattice constant $a_0$ optimized theoretically for $L2_\text{1}$-ordered Fe$_{2}$MnSi. For the stoichiometric composition $x = 1$, the Fermi level ($E_\mathrm{F}$) is situated within the band gap of the down-spin band, i.e., the electronic structure is half-metallic, which is consistent with previous predictions.\cite{Fujii,Hongzhi} We note that a half-metallic electronic structure can be seen even for a nonstoichiometric composition of $x = 0.75$. We define the spin polarization $P$ of the DOS as $[D_\uparrow(E_\mathrm{F})-D_\downarrow(E_\mathrm{F})] / [D_\uparrow(E_\mathrm{F})+D_\downarrow(E_\mathrm{F})]$. Then, the half-metallic electronic structure ($P = +1$) is obtained in the range $0.75 \le x \le 1.5$. With decreasing $x$, the half-metallicity tends to be lost and $P$ begins to decrease, but a high spin polarization of $P =$ 0.9 remains down to $x =$ 0.5. When $x$ is further reduced, $P$ decreases steeply and takes a negative value of $P = - 0.36$ for $x = 0$. It might be worth noting that the magnetic moment obtained for the half-metallic state ($0.75 \le x \le 1.5$) is well explained by the Slater-Pauling behavior of the full-Heusler alloys.\cite{Galanakis}
\begin{figure}[t]
\includegraphics[width=8.5cm]{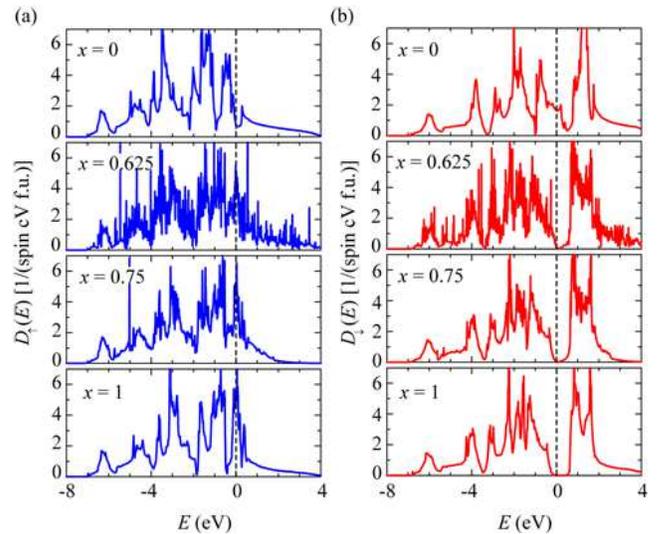}
\caption{(Color online) Densities of states for (a) spin-up ($\uparrow$) and (b) spin-down ($\downarrow$) states of cubic Fe$_{3-x}$Mn$_{x}$Si for $x = 0$, 0.625, 0.75, and 1. The Fermi level is chosen as the origin of energy.}
\end{figure}      

We also investigated theoretically how the electronic structure of Fe$_{3-x}$Mn$_{x}$Si is affected by lattice distortion under biaxial compressive or tensile strain. Assuming (001) stacking and a constant volume of the unit cell, we introduce a tetragonal distortion, $a^2 c = a_0^3$, where $a$ and $c$ are in-plane and out-of-plane lattice constants, respectively. Figure 4 shows a contour plot of $P$ obtained for various $x$ and $c/a$. It is clear that $P$ gradually decreases with increasing lattice distortion. For the half-metallicity of $X$$_{2}$Mn$Z$ Heusler compounds, it is known that the energy level splitting between the $t_{2g}$ and $e_g$ orbitals of the $X$ ions due to the crystal field in cubic symmetry is a crucial factor.\cite{Galanakis} As the crystal lattice deviates from cubic (i.e., $c/a$ deviates from 1), the energy splitting shrinks and then the half-metallicity is lost, causing a decrease in $P$. However, we find that the half-metallicity is stable in a relatively wide range of $x$ and $c/a$ for the Fe$_{3-x}$Mn$_{x}$Si systems considered here. In experiments, although the bulk lattice constant of Fe$_{3-x}$Mn$_{x}$Si ($x \approx 1$) has been reported to be 0.5665 $\sim$ 0.5672 nm,\cite{Yoon,Niculescu,Hongzhi,Zhang} that of our epilayer shown in Fig. 1(b) becomes $\sim$ 0.5654 nm, slightly smaller than the bulk value. It can be considered that the estimated lattice constant of the MBE-grown epilayer includes an effect due to lattice distortion. Hence, it is important for a full description of epitaxial Heusler compounds to understand the influence of lattice distortion on the half-metallicity. We note that the lattice distortion for $x \approx$ 0.6 is quite small, $\sim$ 0.7\%, where the converted $c/a$ is in the range 0.98 $ \le c/a \le 1.02$. Therefore, we can expect that room-temperature ferromagnetic Fe$_{3-x}$Mn$_{x}$Si/Ge epilayers with $x \approx$ 0.6 maintain a high spin polarization of $P \ge$ 0.9, and this material can be used as spin injector and detector in Ge-based spintronic devices. 

Because of interdiffusive solid-phase reactions\cite{Maeda} at the Fe$_{3-x}$Mn$_{x}$Si/Ge interface, we could not utilize high-temperature processes ($\ge$ 500 $^{\circ}$C) to fabricate the magnetic tunnel junctions with Ge or MgO high-quality tunnel barriers. Thus, we cannot estimate the spin polarization of the Fe$_{3-x}$Mn$_{x}$Si epilayers experimentally. To advance the next step for group-IV-semiconductor spintronics, we should further explore a method for the thermal stabilization of the Heusler compounds/Ge interfaces.
\begin{figure}[t]
\includegraphics[width=8.5cm]{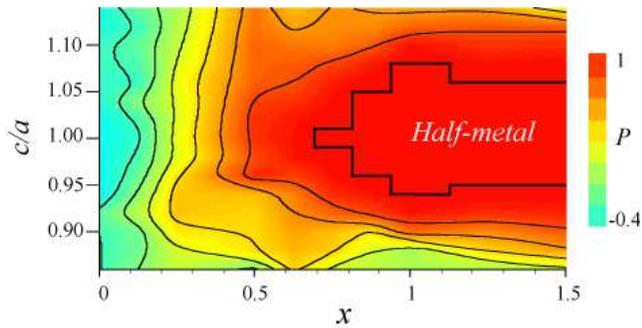}
\caption{(Color online) Contour plot of spin polarization $P$ as functions of $x$ and $c/a$, where the contour lines are drawn at intervals of $|P| =$ 0.2.}
\end{figure} 
 
In summary, we have explored control of ferromagnetism and electronic structure of Heusler compounds Fe$_{3-x}$Mn$_{x}$Si epitaxially grown on a group-IV semiconductor Ge. By tuning the Mn composition $x$ during LT-MBE growth, the saturation magnetization and Curie temperature can be controlled and we can fabricate $L2_\text{1}$-ordered structures, even for nonstoichiometric atomic compositions. We demonstrated that for a composition of $x \approx$ 0.6, the Curie temperature exceeds 300 K. Theoretical calculations suggest that Fe$_{3-x}$Mn$_{x}$Si alloys with nonstoichiometric compositions in the range 0.75 $\le$ $x$ $\le$ 1.5 have a half-metallic electronic structure. The effect of lattice distortion on the spin polarization $P$ was also examined. It is expected that room-temperature ferromagnetic Fe$_{3-x}$Mn$_{x}$Si/Ge epilayers maintain a high spin polarization of $P \ge$ 0.9. 

K.H. and M.M. wish to thank Y. Maeda, T. Sadoh, Y. Nozaki, Y. Terai, and Y. Ando for their helpful discussions, and S. Zaima, K. Matsuyama, M. Nishida, and Y. Kitamoto for providing the use of their facilities. This work was partly supported by a Grant-in-Aid for Scientific Research on Priority Area (No.18063018) from the Ministry of Education, Culture, Sports, Science, and Technology in Japan.  


\end{document}